\documentclass[%
 reprint,
superscriptaddress,
 amsmath,amssymb,aps,
 prl,
]{revtex4-2}

\usepackage{graphicx}
\usepackage{dcolumn}
\usepackage{bm}
\usepackage{amsmath}
\usepackage{xcolor}
\usepackage{changes}
\usepackage{xfrac}
\usepackage{mathtools}
\DeclarePairedDelimiter\abs{\lvert}{\rvert}
\usepackage{upgreek}
\usepackage{gensymb}
\usepackage{appendix}
\usepackage{subfiles}

\begin{document}

\preprint{APS/123-QED}

\title{Tweaking Spectral Topology and Exceptional Points by Nonlinearity \\ in Non-Hermitian Polariton Systems}

\author{Jan Wingenbach}
 \affiliation{Department of Physics and Center for Optoelectronics and Photonics Paderborn (CeOPP), Paderborn University, Warburger Strasse 100, 33098 Paderborn, Germany}

\author{Stefan Schumacher}
\affiliation{Department of Physics and Center for Optoelectronics and Photonics Paderborn (CeOPP), Paderborn University, Warburger Strasse 100, 33098 Paderborn, Germany}%
\affiliation{Institute for Photonic Quantum Systems (PhoQS),
Paderborn University, Warburger Straße 100, 33098 Paderborn, Germany}%
\affiliation{Wyant College of Optical Sciences, University of Arizona, Tucson, AZ 85721, USA}%

\author{Xuekai Ma}
\affiliation{Department of Physics and Center for Optoelectronics and Photonics Paderborn (CeOPP), Paderborn University, Warburger Strasse 100, 33098 Paderborn, Germany}%

\begin{abstract}
Exceptional points (EPs) with their intriguing spectral topology have attracted considerable attention in a broad range of physical systems, with potential sensing applications driving much of the present research in this field. Here we theoretically demonstrate the realization of EPs in a system with significant nonlinearity, a non-equilibrium exciton-polariton condensate. With the possibility to control loss and gain and nonlinearity by optical means, this system allows for a comprehensive analysis of the interplay of nonlinearities (Kerr-type and saturable gain) and non-Hermiticity. Not only do we find that EPs can be intentionally shifted in parameter space by the saturable gain, we also observe intriguing rotations and intersections of Riemann surfaces, and find nonlinearity-enhanced sensing capabilities. Our results are quite general in nature and illustrate the potential of tailoring spectral topology and related phenomena in non-Hermitian systems by nonlinearity.
\end{abstract}

\maketitle
\emph{Introduction} -- Exceptional points (EPs) are singularities in parameter space at which two or more eigenvalues and their corresponding eigenvectors coalesce \citep{Berry2004, Bender2007, Heiss2012, heiss2004exceptional}. Such singularities occur exclusively in non-Hermitian systems which are subject to gain and loss and exhibit non-orthogonal eigenvectors and complex eigenvalues~\citep{kato2013perturbation, PhysRevResearch.1.033182}. Compared to conventional singularities in Hermitian systems, known as Diabolic points, EPs show intriguing properties due to the spectral topology of their Riemann surfaces~\citep{doi:10.1126/science.aar7709, ding2022non}. EPs have been widely investigated in a variety of physical systems such as microwave resonators~\citep{PhysRevLett.90.034101, PhysRevLett.86.787, RevModPhys.87.61}, atomic systems~\citep{PhysRevLett.104.153601}, plasmonic nanostructures~\citep{PhysRevB.94.201103, park2019observation}, optical waveguides~\citep{PhysRevLett.103.093902}, and microresonators~\citep{PhysRevLett.103.134101, chang2014parity, peng2014parity}. Near an EP, a range of counterintuitive phenomena has been reported in optical systems due to the coalescence of the eigenvectors, including loss- and optomechanically-induced transparency~\citep{Zhang:18, lu2018optomechanically}, unidirectional invisibility and reflectivity exceeding unity~\citep{PhysRevLett.103.093902, PhysRevLett.106.213901}. Moreover, sensing enhancement can be realized close to an EP of order $n$ where the frequency splitting scales as the $n^\mathrm{th}$-root of the perturbation~\citep{Chen2017, Hodaei2017, Wiersig:20, hodaei2017enhanced, mandal2021symmetry}. This makes EPs promising candidates for a new generation of sensors, with the potential to outperform their Hermitian counterparts~\citep{PhysRevLett.112.203901, PhysRevA.93.033809, wiersig2020prospects}. 

Recently, there has been growing interest in the interplay of nonlinearity and non-Hermiticity. For instance, the influence of nonlinear effects on the $\mathcal{PT}-$symmetry in lasing systems has been studied~\citep{Ge2016, Teimourpour2017, PhysRevA.92.063807, PhysRevA.96.043836}. EP-based sensors with saturable gain~\citep{PhysRevApplied.18.054059}, saturable-gain induced energy shift~\citep{PhysRevA.103.043510, el2023tracking}, and the encircling of EPs in bistable domains~\citep{Wang:19} were investigated. To systematically study the interplay of nonlinear and non-Hermitian physics, systems with variable nonlinearity and controllable gain and loss are required. An example of such systems are exciton polariton condensates, in which rich nonlinear physics have been reported~\citep{PhysRevB.105.245302, Ma2020, PhysRevLett.121.227404, li2022manipulating, Ballarini2013, Daskalakis2014}. Exciton polaritons are hybrid light-matter quasiparticles that form due to strong light-matter coupling, pairing finite lifetimes on a picosecond scale, and thus non-Hermiticity, with strong nonlinearity from polariton-polariton interactions \citep{deng2002condensation,kasprzak2006bose}. For nonresonant optical excitation, spontaneous macroscopic coherence can form, known as polariton condensation~\citep{Kasprzak2006, doi:10.1126/science.1074464}. 
In that case, the significant interaction of the condensate and the exciton reservoir induces a repulsive potential energy landscape, enabling optical trapping~\citep{schmutzler2015all} as well as the optical control of the polariton condensate~\cite{Ma2020}. The nonlinearity inside the polariton condensate can be tuned with static electric fields~\citep{PhysRevLett.121.037401, De2023, zhai2022electrically}. With their non-equilibrium nature and the possibility of both structural and optical control, polariton condensates offer a natural playground for the study of non-Hermitian physics \citep{gao2018chiral, gao2015observation, Li2022}. However, the interplay of nonlinearity and non-Hermiticity has previously not been systematically explored or exploited.

In the present work, we investigate the manipulation of EPs and surrounding Riemann surfaces through the nonlinearities in the system. We find that the saturable gain shifts the EP in parameter space, leading to a variation in the mode coupling requirement for the observation of the EP. The repulsive polariton-polariton nonlinearity (akin to a Kerr-type nonlinearity) can induce not only an energy blueshift but also a simultaneous rotation of the Riemann surface and movement of the EP. We show that this also applies to higher-order EPs where the Riemann surfaces can show complex intersection patterns. With potential applications in mind, we further demonstrate that sensing sensitivity near the EP can be significantly enhanced by nonlinearity. 

\begin{figure}[t]
  \centering
   \includegraphics[width=\columnwidth]{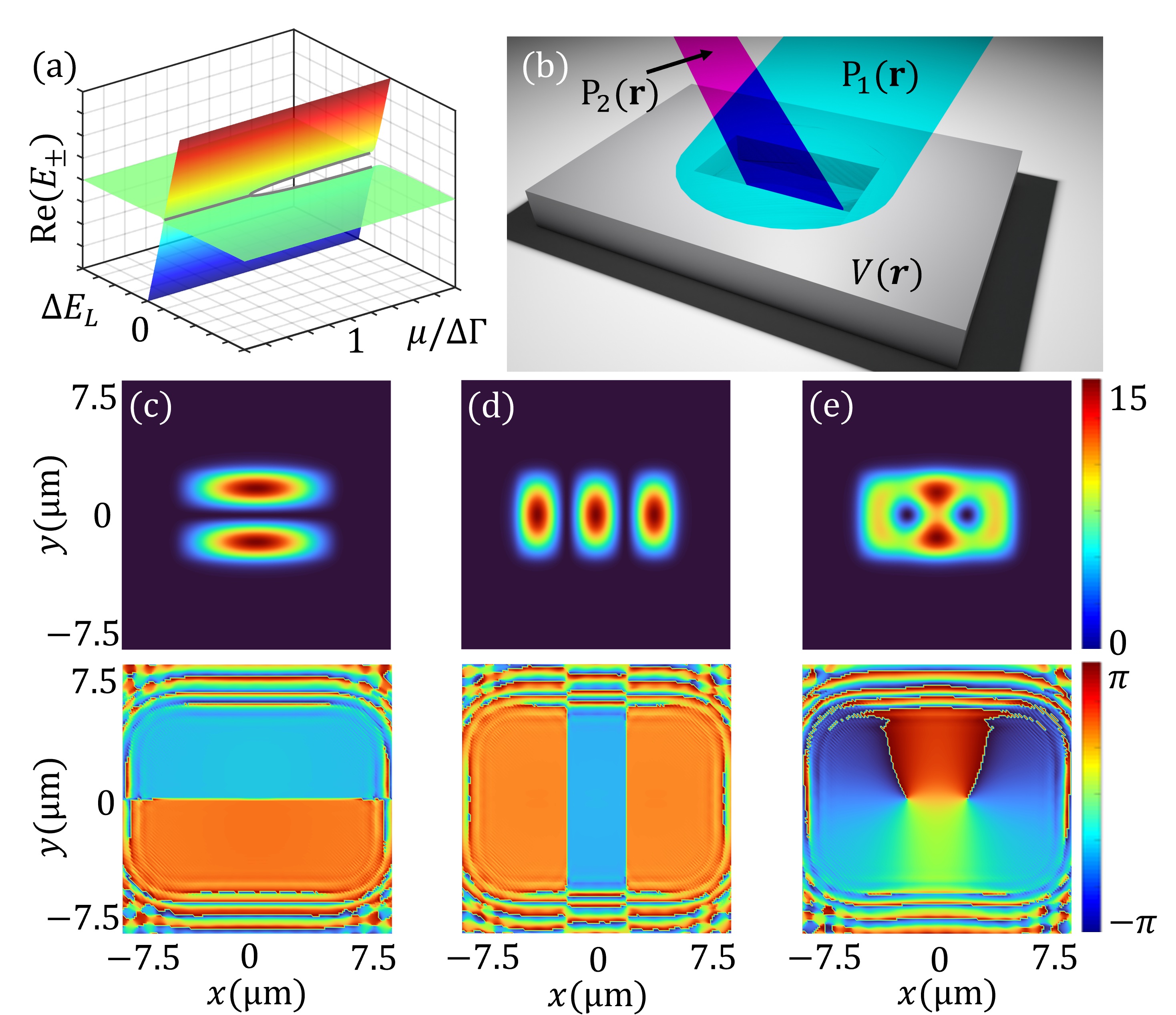}
  \caption{\textbf{System and excitation setup.} (a) Sketch of Riemann surface (real part only) of a two-level non-Hermitian system containing an EP for varying level separation $\Delta E_L=E_\alpha-E_\beta$ and varying relative level coupling $\mu/\Delta\Gamma$. (b) Scheme for realization with polariton condensates, including rectangular external potenial, $V(\textbf{r})$, and excitation with two nonresonant pump beams, $P_1(\textbf{r})$ and $P_2(\textbf{r})$. Polariton density (upper row; in $\mu m^{-2}$) and phase (lower row) of the (c) third and (d) fourth mode of the potential $V(\textbf{r})$, underlying the realization of the EP. (e) Vortex-antivortex mode at the EP. Further details are given in the text.}
  \label{fig:1}
\end{figure}

\emph{Theoretical~Model} -- To study the dynamics of polariton condensates under nonresonant excitation, we use the dissipative Gross-Pitaevskii (GP) model \cite{PhysRevLett.99.140402, PhysRevB.105.245302}. Within this framework, the dynamics of the coherent polariton field $\psi=\psi(\textbf{r},t)$ and the density of the exciton reservoir $n=n(\textbf{r},t)$ are
\begin{eqnarray}
\label{GP_psi}
    \mathrm{i}\hbar\frac{\partial}{\partial t}\psi = \biggl(H_0(\textbf{r})+g_\mathrm{c}|\psi|^2+g_\mathrm{r}n+\frac{\mathrm{i}\hbar}{2}[Rn-\gamma_\mathrm{c}]\biggr)\psi\,,
\end{eqnarray}
\begin{equation}
\label{GP_n}
    \frac{\partial{n}}{\partial{t}} = P(\textbf{r}) - (\gamma_\mathrm{r}+R|\psi|^2)n\,.\nonumber
\end{equation}
The linear operator $H_0(\textbf{r})=-\frac{\hbar^2}{2m_\mathrm{eff}}\nabla^2+V(\textbf{r})$ describes the kinetic energy and the external potential $V(\textbf{r})$ which confines the condensate spatially. $m_\mathrm{eff}=10^{-4}m_e$ is the effective polariton mass with $m_e$ being the free electron mass. $g_c$ is the strength of the polariton-polariton interaction (akin to a Kerr-type nonlinearity), while $g_r$ is the strength of the polariton-reservoir interaction. $\gamma_c=0.3~\mathrm{ps^{-1}}$ is the loss constant of the condensate in quasi-mode approximation \cite{Carcamo:20}, counteracted by the stimulated in-scattering with $R=0.01~\mathrm{ps^{-1}\upmu m^{2}}$. $\gamma_r=0.45~\mathrm{ps^{-1}}$ is the loss constant of the reservoir, supplemented by the incoherent pump $P(\textbf{r})$. For stationary excitation and solution we obtain $n(\textbf{r})=\frac{P(\textbf{r})}{\gamma_r+R|\psi(\textbf{r})|^2}$; the term $R|\psi(\textbf{r})|^2$ holds the saturable gain.

A system containing an EP of two coalescing eigenvectors is approximated by a two-level effective Hamiltonian,
\begin{equation}
    H_\mathrm{EP2}=\begin{pmatrix}E_\upalpha+i\Gamma_\upalpha & \sfrac{\mu}{2} \\ \sfrac{\mu}{2} &E_\upbeta+i\Gamma_\upbeta\end{pmatrix}.
    \label{eq:EP2}
\end{equation}
Here $E_\upalpha$ and $E_\upbeta$ are the energies of mode $\upalpha$ and $\upbeta$. The gain and loss are included into $\Gamma_\upalpha$ and $\Gamma_\upbeta$, $\mu$ characterizes the coupling strength of the two modes. The eigenvalues of the Hamiltonian \eqref{eq:EP2} read $E_\pm=\sfrac{\left[E_\upalpha+E_\upbeta+i(\Gamma_\upalpha+\Gamma_\upbeta)\right]}{2} \pm\sfrac{\sqrt{|\mu|^2+[E_\upalpha-E_\upbeta+i(\Gamma_\upalpha-\Gamma_\upbeta)]^2}}{2}$. The second term describes the energy splitting on one side of the EP which shows the characteristic square-root-dependency. Plotting the eigenvalues in dependence of the energy difference $\Delta E_\mathrm{L}=E_\upalpha-E_\upbeta$ and the loss and gain difference $\Delta\Gamma=\Gamma_\upalpha-\Gamma_\upbeta$ visualizes the Riemann surface of the system as shown in Fig.~\ref{fig:1}(a) for the real part. The EP is localized at $\Delta E_\mathrm{L}=0$ and $\mu=\Delta\Gamma$.

To study the dynamics of two coupled modes $\psi_{\upalpha,\upbeta}$ of a polariton condensate in the vicinity of an EP, we extend the effective non-Hermitian Hamiltonian \eqref{eq:EP2} by including the nonlinearity of the GP model. 
Consequently, the two-level non-Hermitian Hamiltonian and the effective nonlinear Schrödinger equation (NSE) read
\begin{equation}      
        i\hbar\frac{\partial}{\partial t}\begin{pmatrix}\psi_{\upalpha}\\\psi_{\upbeta}\end{pmatrix} = \begin{pmatrix}H(\psi_{\upalpha}) & \sfrac{\mu}{2}\\ \sfrac{\mu}{2} & H(\psi_{\upbeta}) \end{pmatrix}\begin{pmatrix}\psi_{\upalpha}\\\psi_{\upbeta}\end{pmatrix},      
        \label{eq:1}
\end{equation}
where $H(\psi)=H_0(\textbf{r})+g_\mathrm{c}|\psi|^2+g_\mathrm{r}n+\frac{\mathrm{i}\hbar}{2}[Rn-\gamma_\mathrm{c}]$.
In the further course of this work, the eigenenergies of the system are denoted by $E_{\upalpha,\upbeta}$ and their difference by $\Delta E_\mathrm{L}$. The blueshift-induced correction to this difference is defined as $\Delta E_\mathrm{NL}$, such that the energy difference in the nonlinear regime is $\Delta\mathcal{E}=\Delta E_\mathrm{L}+\Delta E_\mathrm{NL}=\mathcal{E}_{\upalpha}-\mathcal{E}_{\upbeta}$. In this work, we investigate the third, $\upalpha=3$, and fourth, $\upbeta=4$, mode as shown in Fig.~\ref{fig:1}(c,d) of a rectangular external potential with width of $5~\mathrm{\upmu m}$, length of $8.9~\mathrm{\upmu m}$, and depth of $2~\mathrm{meV}$, as sketched in Fig.~\ref{fig:1}(b). These two modes are chosen as their energy is close to each other and their degeneracy can be realized in experiments~\cite{gao2015observation}. 
We use a nonresonant pump $P(\textbf{r})$ which consists of a broad flat top pump $P_1(\textbf{r})$ [blue light cone in Fig. \ref{fig:1}(b)] with diameter of $12~\mathrm{\upmu m}$ and intensity of $I_1=16~\mathrm{ps^{-1} \upmu m^{-2}}$. An additional elliptical pump $P_2(\textbf{r})$ [pink light cone in Fig.~\ref{fig:1}(b)] with height of $2~\mathrm{\upmu m}$ and width of $5~\mathrm{\upmu m}$ is used to tune the mode energy difference $\Delta\mathcal{E}$. Mathematical expressions for the pump profiles are given in the SM.
The total energies $\mathcal{E}_{\upalpha,\upbeta}$ of the two modes are obtained by solving the GP equation, Eq.~(\ref{GP_psi}), by the Dormand-Prince method and time-domain Fourier transformation for the respective steady states [Fig.~\ref{fig:1}(c,d)]. Their scalar gain and loss rates can be approximated by $\Gamma_{\upalpha,\upbeta}\approx N^{-1}\frac{\hbar}{2}\int \left [ Rn_{\upalpha,\upbeta}(\textbf{r})-\gamma_c \right ] |\psi_{\upalpha,\upbeta}(\textbf{r})|^2\mathrm{d^2}\textbf{r}$ with $N^{-1}=\int|\psi_{\upalpha}(\textbf{r})|^2\mathrm{d^2}\textbf{r}$~\citep{gao2018chiral}. The complex eigenvalues $E_\pm$ of the corresponding effective two-level system, Eq.~(\ref{eq:EP2}), can then be determined for a given $\Delta\mathcal{E}=\mathcal{E}_{\upalpha}-\mathcal{E}_{\upbeta}$ and as a function of the mode coupling strength $\mu$ with the nonlinear energies as input for $E_{\upalpha,\upbeta}$. For the coupled two-level NSE, Eq.~\eqref{eq:1}, with $\mu=\Delta\Gamma$ close to the degeneracy $\Delta\mathcal{E}=0$, the two modes coalesce into the superposition mode in which a vortex-antivortex pair forms [see Fig.~\ref{fig:1}(e)]. By encircling the EP, supported by $P_2$, the two vortices swap their topological charge, which is a typical property of a topological Riemann surface \cite{gao2018chiral} (details of encircling in SM Fig.~S2).

\begin{figure}[tb]
  \centering
   \includegraphics[width=1.0\columnwidth]{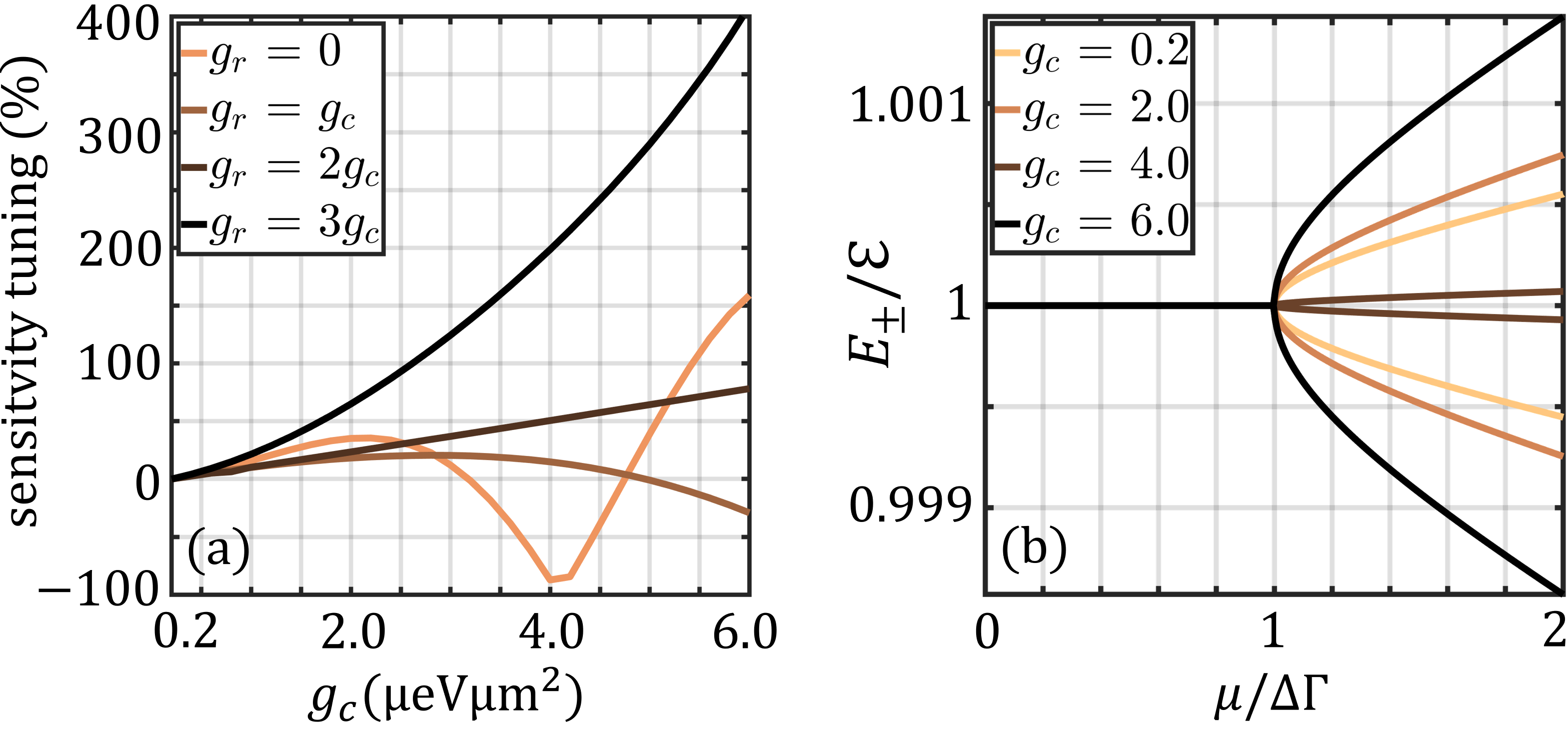}
  \caption{\textbf{Nonlinearity-induced enhancement of sensing at EP.} (a) Deviation of the eigenvalue splitting (real part) from the linear case at $\sfrac{\mu}{\Delta\Gamma}=2$, close to the EP as a function of the polariton-polariton interaction $g_\mathrm{c}$ and for different polariton-reservoir interactions  $g_\mathrm{r}$. The values of the splitting deviation and the corresponding increased or decreased EP sensitivity are given as a percentage of the energy splitting in the linear case. (b) Eigenvalues $E_{\pm}$ along the mode degeneracy line $\Delta\mathcal{E}=0$ for different interaction strengths $g_\mathrm{c}$ and $g_\mathrm{r}=0$, illustrating the change in the energy splitting. Eigenvalues are renormalized to the total mode energy $\mathcal{E}$ at the EP.}
  \label{fig:2}
\end{figure}

\emph{Sensing~Enhancement} -- We first focus on the influence of nonlinearities on the complex eigenvalues along the degeneracy of real-valued mode energies $\Delta\mathcal{E}=0$. The eigenvalues $E_\pm$ of the 2D system are determined as a function of $\mu$, by evaluating the eigenvalues according to the method introduced above. Exposed to a saturable gain, which in our case is expressed by $R|\psi(\textbf{r})|^2$, the gain difference of the modes in a nonlinear system depends on their intensity. The coupling of the modes in our multistable system is also influenced by the strength of the polariton-polariton interaction. Thus, the necessary coupling strength $\mu$ to reach the EP in such a nonlinear system is changed by the intensity of the modes or in this case the two-dimensional polariton density. Hence, we observe a shift of the EP along the $\Delta\Gamma$ or $\mu$ axis as a function of the nonlinear interaction strength (see SM Fig.~S1). A similar effect was recently observed in Ref.~\citep{el2023tracking} in a lasing system.
For stationary solutions with $|\psi_{\upalpha,\upbeta}|^2=\mathrm{max}\left(|\psi_{\upalpha,\upbeta}(\textbf{r})|^2\right)$ and $n_{\upalpha,\upbeta}=\mathrm{max}\left( n_{\upalpha,\upbeta}(\textbf{r})\right)$, the eigenvalues of Eq.~(\ref{eq:EP2}) with respect to the nonlinear terms of Eq.~(\ref{GP_psi}) reads
\begin{eqnarray}
        E_{\pm,\mathrm{NL}} &\approx&\frac{\left(E_\upalpha +E_\upbeta+g_c|\psi_{\mathrm{tot}}|^2+g_r n_{\mathrm{tot}}+i\left(\Gamma_\upalpha+\Gamma_\upbeta\right)\right)}{2} \nonumber\\
        & \pm & \frac{\sqrt{|\mu|^2+\left(\Delta E_\mathrm{L}+\Delta E_\mathrm{NL}+i\Delta\Gamma\right)^2}}{2}\;.
        \label{eq:2}
\end{eqnarray}
Here the polariton- and reservoir-polariton interactions induce an energy blueshift on the Riemann surface proportional to the total polariton $|\psi_{\mathrm{tot}}|^2=|\psi_\upalpha|^2+|\psi_\upbeta|^2$ and reservoir density $n_{\mathrm{tot}}=n_\upalpha+n_\upbeta$. $\Delta E_{\mathrm{NL}}=g_c\Delta|\psi|^2+g_r\Delta n$ is characterized by the density difference $\Delta|\psi|^2=|\psi_\upalpha|^2-|\psi_\upbeta|^2$ and $\Delta n=n_\upalpha-n_\upbeta$. While the two modes coalesce at the EP with $\Delta|\psi|^2=0$ and $\Delta n=0$, tuning the system away from the EP also lifts the coalescence of the two modes, resulting in nonvanishing contributions of $\Delta|\psi|^2$ and $\Delta n$. Depending on the exact shape of the two modes, this can lead to an increase ($\Delta E_\mathrm{NL}>0$) or decrease ($\Delta E_\mathrm{NL}<0$) of the eigenvalue splitting. The sensitivity change $\kappa$ for a given ratio $\sfrac{\mu}{\Delta\Gamma}$ can be derived as discussed in the SM.

After solving Eq.~(\ref{eq:1}), the eigenvalues of Eq.~(\ref{eq:EP2}) are calculated. Then their splitting $\kappa$ is determined for $\sfrac{\mu}{\Delta \Gamma}=2$ and the percentage deviation from the splitting at the reference value is calculated. The results are plotted in Fig. \ref{fig:2}(a) as a function of $g_c$ for different $g_r$. For $g_r=0$, the sensitivity of the EP can increase by up to $150\%$ and decrease to about $10\%$. For comparison, the resulting EPs are shown renormalized by the respective mode energy at $\Delta\mathcal{E}=0$ and their gain differences [see Fig.~\ref{fig:2}(b)]. For $g_r\neq0$, the oscillation of the sensitivity decreases. For values of $g_r\geq2g_c$, the sensitivity increases monotonically as the reservoir interaction dominates. It is worth pointing out that $g_c$ depends on the detuning of the exciton and photon states and the choice of material~\citep{PhysRevB.100.035306} and can be controlled during sample preparation or by applying a static electric field~\citep{PhysRevLett.121.037401, De2023, zhai2022electrically}.

\begin{figure}[t]
  \centering
   \includegraphics[width=\columnwidth]{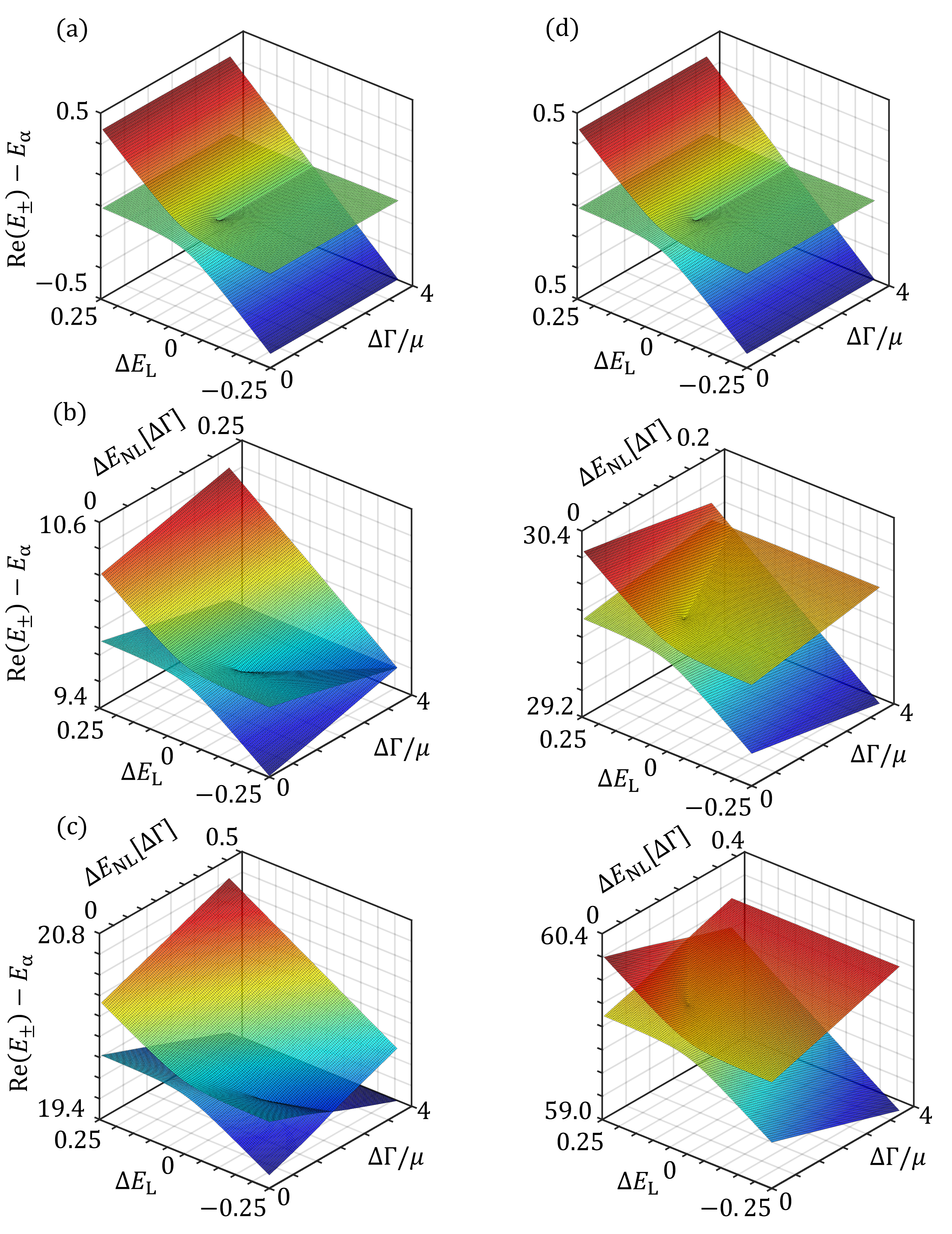}
  \caption{\textbf{Nonlinearity-induced rotation of Riemann surfaces.} Riemann surfaces (real parts) resulting from the eigenvalue calculations, Eq.~(\ref{eq:2}), for (a-c) $g_\mathrm{r}=0$ and (d-f) $g_\mathrm{r}=2g_\mathrm{c}$ for different polariton-polariton interaction: (a,d) $g_\mathrm{c}=0~\mathrm{\upmu eV\upmu m^2}$, (b,e) $g_\mathrm{c}=3~\mathrm{\upmu eV\upmu m^2}$ and (c,f) $g_\mathrm{c}=6~\mathrm{\upmu eV\upmu m^2}$. The eigenvalues are depicted against the energy difference $\Delta E_\mathrm{L}$ and relative gain and loss difference $\Delta\Gamma/\mu$. For $g_\mathrm{c}\neq0$ the nonlinear correction $\Delta E_\mathrm{NL}$ to the energy difference is plotted with $\Delta\Gamma$. Energies are given in $\mathrm{\upmu eV}$.}
  \label{fig:3}  
\end{figure}

\emph{Riemann Surface Rotation} -- We extend our investigation from the bifurcation of the eigenvalues at the EP along the trace of degeneracy to the entire Riemann surface in the nonlinear regime. Here, we determine the eigenvalues by using Eq.~(\ref{eq:2}) in dependence of $\Delta E_\mathrm{L}$ and $\Delta\Gamma$. Spatially varying quantities are replaced by scalar approximations. Thus, the flat top pump $P$ is approximated by its peak intensity $I_1$ and the mode densities by their average values $|\overline{\psi}|^2\approx3~\mathrm{\upmu m^{-2}}$ inside the external potential. $E_\upalpha$ is fixed to the eigenenergy of the dipole mode as shown in Fig.~\ref{fig:1}(c), while $E_\upbeta$ is varied close to this value. The resulting energy difference is linked to the ratio of the length and width of the rectangular potential~\cite{gao2015observation}. The mode coupling is extracted from the calculations performed in the above section. To construct the Riemann surface, the gain and loss difference is varied independently. To this end, we assume that the total density $|\psi_\mathrm{tot}|^2$ is resilient to small variations of $\Delta\Gamma$. The remaining parameters are taken from the GP model calculations. The nonlinear correction to the energy difference $\Delta E_\mathrm{NL}=g_c\Delta|\psi|^2$ is determined by deriving an expression for the polariton density contrast resulting from a gain and loss difference $\Delta\Gamma$, which reads
\begin{equation}
     \Delta|\psi|^2 = \frac{2\Delta\Gamma}{\left[\frac{\hbar PR}{\gamma_\mathrm{r}}-\hbar\gamma_\mathrm{c}\right]+\sum_{m=2}^{\infty}\frac{\hbar PR^m}{\gamma_\mathrm{r}^m}\left(-|\psi_\mathrm{tot}|^2\right)^{m-1}}.
     \label{eq:4}
\end{equation}
The derivation of this expression is found in the SM. The terms within the sum in the denominator follow from the Taylor series of the average gain and loss difference $\overline{\Gamma}=\frac{\hbar}{2}\left[\frac{RP}{\gamma_\mathrm{r}+R|\overline{\psi}|^2}-\gamma_\mathrm{c}\right]|\overline{\psi}|^2$, including the saturable gain. For the following study, we consider terms up to $m=6$. Note that the density difference $\Delta|\psi|^2$, which mainly depends on $\Delta\Gamma$, has to be in a reasonable range for convergence. The influence of the Kerr-type nonlinearity $|\psi|^2\psi$ and the satarable gain $R|\psi|^2$ on the Riemann surface is illustrated in Fig.~\ref{fig:3}. Remarkably, the Kerr-type nonlinearity causes the trace of $\Delta\mathcal{E}=0$ not to be parallel to the $\Delta\Gamma$ axis, which leads to rotation of the Riemann surface. In the presence of the saturable gain the rotation angle becomes density dependent and can thus be increased even further [see Fig.~\ref{fig:3}(a-c)]; the nonlinearity-induced blueshift leads to a tilting of the $\Delta\mathcal{E}=0$ trace and movement of the EP. 
In the particular case of the polariton condensate, the polariton-reservoir interaction $g_\mathrm{r}$ can induce an additional blueshift and thus a rotation of the Riemann surface as shown in Fig.~\ref{fig:3}(d-f). The relation between the total reservoir density and the reservoir difference derived from Eq.~(\ref{eq:4}) is discussed in detail in the SM. It shows that the Riemann surface can even be rotated into the opposite direction in virtue of the polariton-reservoir interaction. In polariton condensates the pump intensity can control both  nonlinear effects by tuning the density of the polariton condensate, which can provide an efficient method to actively tweak the spectral topology and the corresponding EP, since in this case no other system parameters need to be varied.

\begin{figure}[t]
  \centering
   \includegraphics[width=1.0\columnwidth]{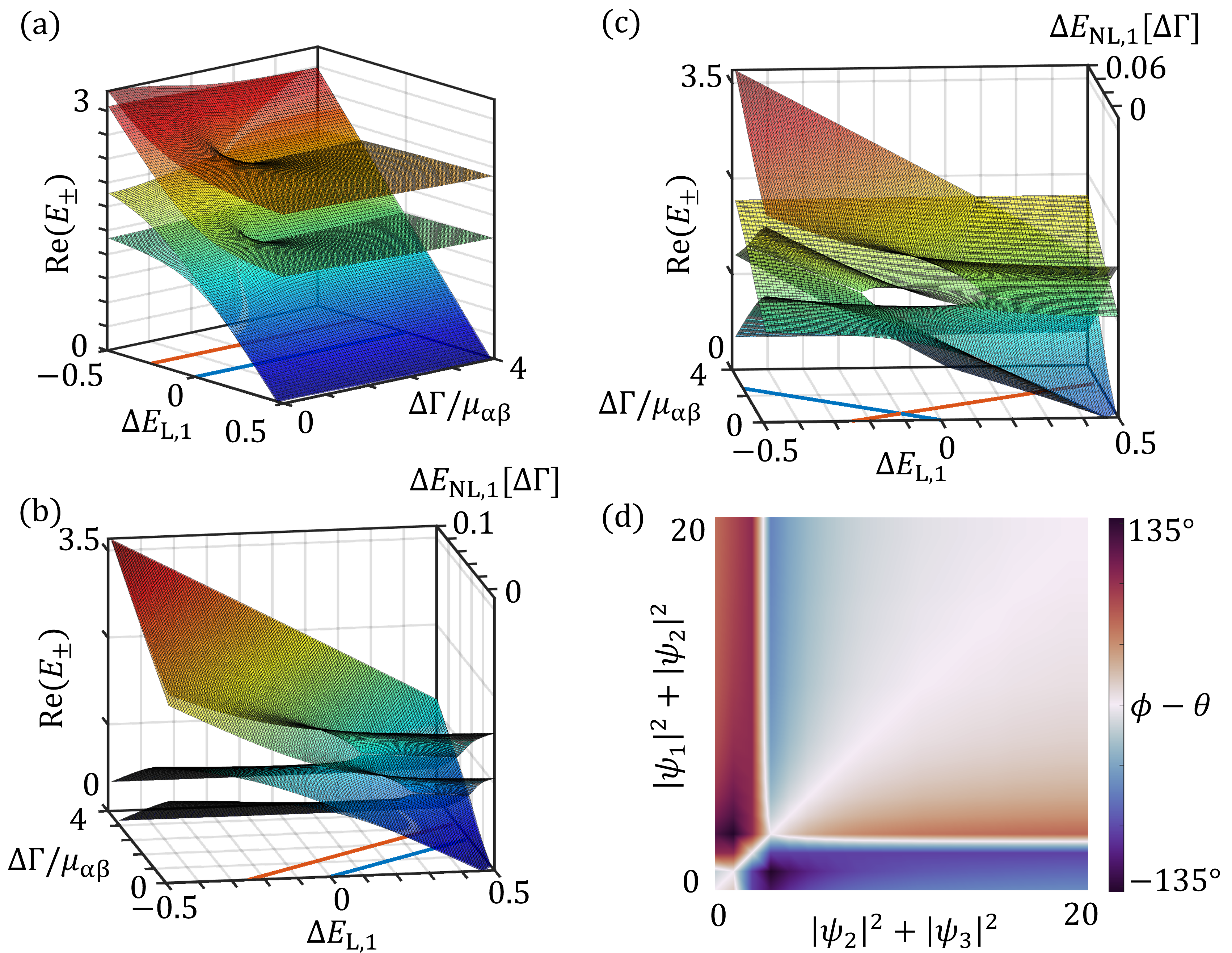}
  \caption{\textbf{Rotation of the Riemann surfaces with higher-order EPs.} Riemann surfaces (real parts) containing two EPs (a) in the linear regime and (b-c) under the impact of the polariton-polariton interaction $g_\mathrm{c}=6~\mathrm{\upmu eV\upmu m^2}$ and the saturable gain. In (b) the total densities of the modes forming the EPs are identical, whereas in (c) the total densities inside each EP are distinct. The eigenvalues are depicted against the eigenenergie difference $E_\mathrm{L}$ and the gain and loss difference $\Delta\Gamma$. For $g_\mathrm{c}\neq0$ the nonlinear correction to the energy difference is plotted with $\Delta\Gamma$. The lines at the bottom of the surfaceplot illustrate the orientation of the $\Delta\mathcal{E}=0$ trace to visualize the rotation of the Riemann surfaces. (d) Deflection angle of the two projected lines as a function of the total densities of the adjacent modes for a nonlinearity of $g_\mathrm{c}=6~\mathrm{\upmu eV\upmu m^2}$. Energies are given in $\mathrm{meV}$.}
  \label{fig:4}
\end{figure}

From the rotation of the Riemann surface, it is worth asking  whether more complicated higher-order EPs and  multiple layer Riemann surfaces can also be manipulated. Here, we consider a parametrized and general three-level non-Hermitian Hamiltonian (with no direct relation to the specific modes in Fig.~1), as motivated for a linear system in Ref.~\citep{demange2011signatures}, which reads
\begin{equation}
H_{\mathrm{EP3}}=\begin{pmatrix}\mathcal{E}_\upalpha+i\Gamma_\upalpha & \sfrac{\mu_{\upalpha\upbeta}}{2} & 0\\ \sfrac{\mu_{\upalpha\upbeta}}{2} &\mathcal{E}_\upbeta+i\Gamma_\upbeta & \sfrac{\mu_{\upbeta\upgamma}}{2} \\ 0 & \sfrac{\mu_{\upbeta\upgamma}}{2} & \mathcal{E}_\upgamma+i\Gamma_\upgamma\end{pmatrix}.
    \label{eq:EP3}
\end{equation}
$\mathcal{E}=E+g_\mathrm{c}|\psi|^2$ denotes the mode energy with nonlinearity. The mode energies $E_\upalpha$ and $E_\upbeta$ and their coupling $\mu_{\upalpha\upbeta}$ refer to the values used above, while $E_\upgamma$ is set to $1.75~\mathrm{meV}$ and $\mu_{\upbeta\upgamma}=0.75\mu_{\upalpha\upbeta}$, providing a clear picture of the resulting Riemann surfaces. In Fig.~\ref{fig:4}(a-c), the Riemann surfaces are shown for different nonlinear contributions. Here, the traces $\Delta\mathcal{E}_{1,2}=0$ are projected onto the bottom of the surface plots to indicate the rotation of the Riemann surfaces. Remarkably, the Riemann surfaces can be rotated as a whole with the two projected lines being parallel [Fig.~\ref{fig:4}(a,b)]. Due to the saturable gain, however, the two Riemann surfaces can also be moved towards opposite directions with the two lines crossing each other [Fig.~\ref{fig:4}(c)], leading to complicated intersection behavior of the Riemann surfaces. Figure~\ref{fig:4}(d) shows the displacement angle difference of the two projected lines in Fig. \ref{fig:4}(a-c) as a function of the total densities of the two adjacent modes. It can be seen that if the saturable gain of the system is negligible or is identical for both EPs, the two lines are not deflected against each other, whereas they are deflected otherwise. It can also be inferred that if the energy blueshift is significant for one or two modes, it may be possible to observe  coalescence of the two EPs, consequently leading to a phase transition in the spectral topology.

\emph{Conclusion} -- We have investigated the influence of nonlinearity on non-Hermitian spectral topology in microcavity polariton condensates. The nonlinearities  lead to a significant change of the eigenvalue splitting near the EP as a function of density difference, leading to a clear nonlinearity-induced sensing capability enhancement. The Kerr-type nonlinearity of the condensate (from polariton-polariton interactions) and the saturable gain (from polariton-reservoir interactions) lead to a rotation of the Riemann surfaces and shift the EP in parameter space. This can give rise to a complex intersection patterns of Riemann surfaces for higher-order EPs. In nonlinear mode control this offers interesting insights for encircling higher-order EPs and for phase transitions at EPs. These results are generic enough to be applied to other non-Hermitian systems with similar nonlinearities such as in nonlinear optics and atomic systems.

\begin{acknowledgments}
\emph{Acknowledgments} -- This work was supported by the Deutsche Forschungsgemeinschaft (DFG) through grants No. 467358803, No. 519608013, and the transregional collaborative research center TRR142/3-2022 (231447078, project A04) and by the Paderborn Center for Parallel Computing, PC$^2$. We thank Tingge Gao for fruitful discussions.
\end{acknowledgments}

\providecommand{\noopsort}[1]{}\providecommand{\singleletter}[1]{#1}%

\cleardoublepage


\section{Supplemental Material}
\renewcommand{\thefigure}{S\arabic{figure}}
\setcounter{figure}{0} 
\setcounter{equation}{0} 
The supplementary material is structured as follows: In the first part, the EP shift along the $\Delta\Gamma$ or $\mu$ axis is presented and the expression for the sensitvity change $\kappa$ in the vicinity of the EP is introduced. In the second part, the encircling of the EP and the resulting vortex-antivortex switching process are discussed. In the last part, the derivation of the relevant equations for Riemann surface rotation is explained.\\
\emph{EP-Shift and Sensitivity tuning} -- Under saturable gain, the gain difference of the modes in a nonlinear system depends on their intensity. Since the coupling of the modes in the multistable system is also affected by the strength of the polariton-polariton interaction, the necessary coupling strength $\mu$ to achieve EP in such a nonlinear system depends on the intensity of the modes or, in this case, on the two-dimensional polariton density. Hence, we observe a shift of the EP along the $\Delta\Gamma$ or $\mu$ axis as a function of the nonlinear interaction strength; see Fig. \ref{fig:sm1}.
\begin{figure}[htb]
  \centering
   \includegraphics[width=1.0\columnwidth]{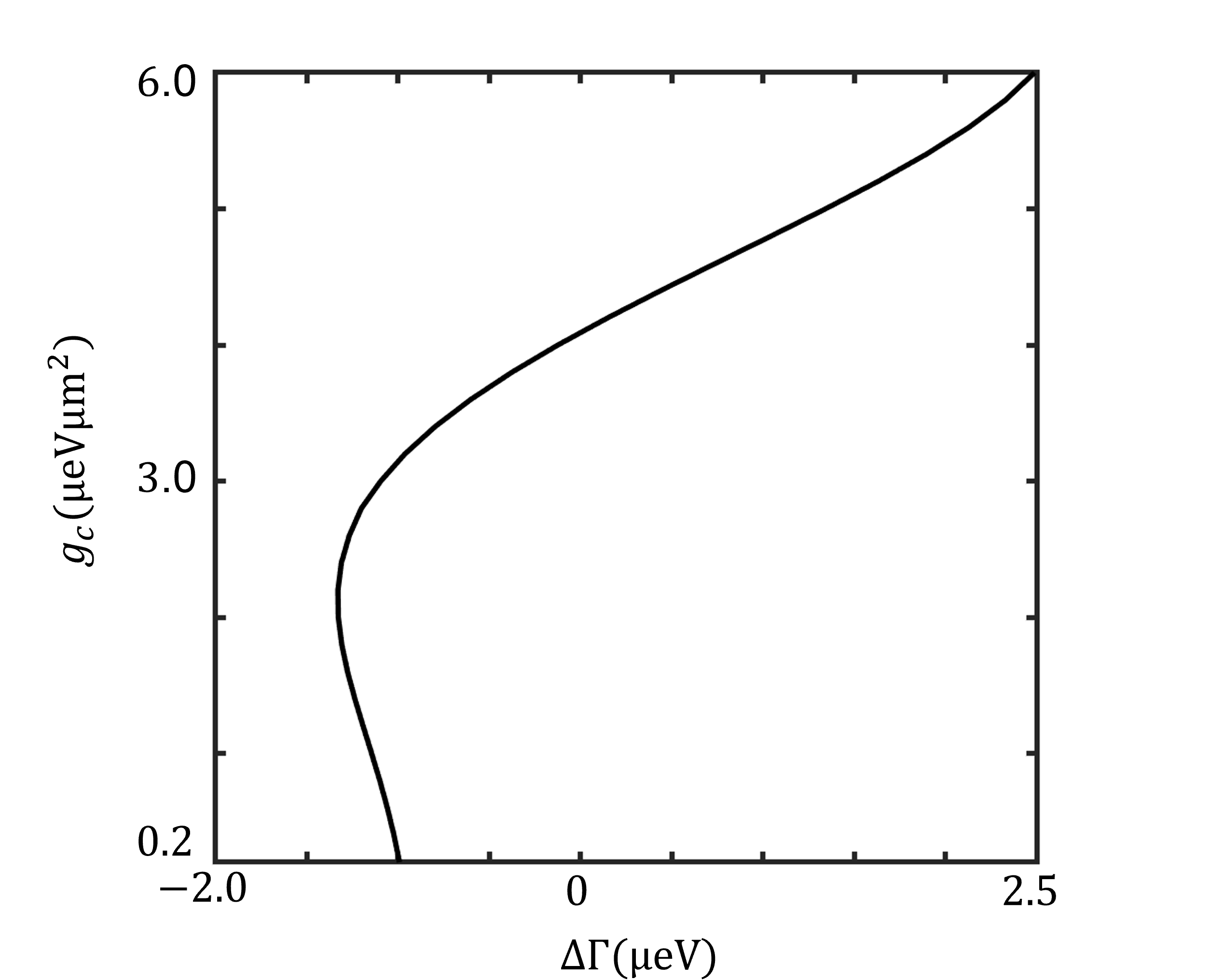}
  \caption{\textbf{Nonlinearity-induced shift of the EP along the $\Delta\Gamma$-axis in the parameter space.} Inverted plot of the EP-position ($x$-axis) shifting under variation of the $g_\mathrm{c}$-induced nonlinearity.}
  \label{fig:sm1}
\end{figure}
Here, the EP position in parameter space is shown on the $x$-axis and the present polariton-polariton interaction strength is plotted on the $y$-axis. Thus it can be seen, that gain and loss difference of the two modes does not only in- or decreases but can also switch sign under increasing Kerr nonlinearity.\\
In the vicinity of an EP of a system subject to a Kerr-like nonlinearity the sensitivity change $\kappa$ for a given ratio $\sfrac{\mu}{\Delta\Gamma}$ can be derived from Eq.~(\ref{eq:2}) for $\Delta E_\mathrm{L}\approx0$ and reads
\begin{eqnarray}
        \kappa=&\pm&\frac{\sqrt{\abs*{\frac{\mu}{\Delta\Gamma}}^2+\left(g_c\Delta|\psi|^2+g_r\Delta n+i\Delta\Gamma\right)^2}}{2\mathcal{E}}\nonumber\\
        &\mp&\frac{\sqrt{\abs*{\frac{\mu}{\Delta\Gamma'}}^2-\Delta{\Gamma'}^2}}{2\mathcal{E'}}\;.
\label{eq:3}
\end{eqnarray}
Here the $\pm$ and $\mp$ signs describe the eigenvalue splitting in both directions. The first term describes the eigenvalue splitting under the influence of the nonlinear interactions $g_c$ and $g_r$, while the second term describes the eigenvalue splitting in the linear case. 
The resulting EPs are superimposed by renormalizing the two scenarios with the respective mode energy at $\Delta\mathcal{E}=0$ and their gain differences. Since the multistability of modes is required in this system, the eigenvalue bifurcation under a weak nonlinearity of $g_c=0.2~\mathrm{\upmu eV\upmu m^2}$ is chosen as a reference value for the sensitivity tuning.\\

\begin{widetext}
\begin{figure*}[htb]
  \centering
   \includegraphics[width=2.0\columnwidth]{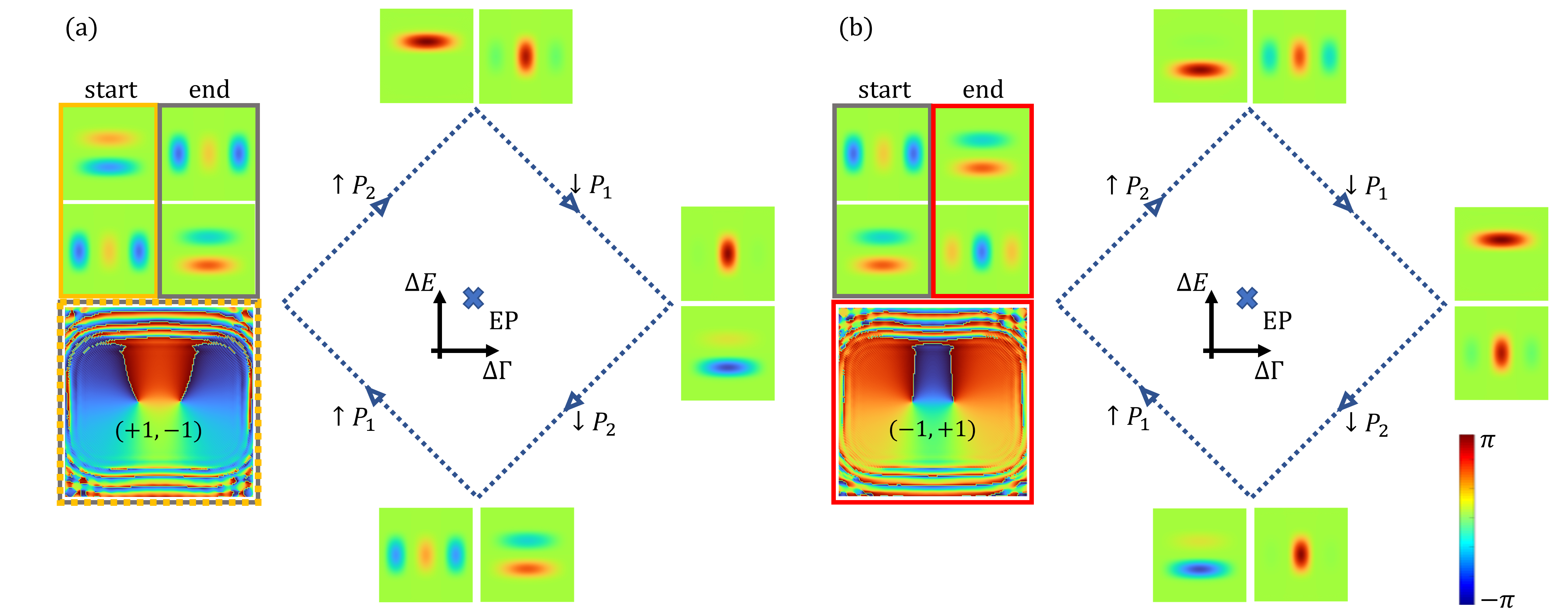}
  \caption{\textbf{All-optical vortex-antivortex switching under EP encircling.} Illustration of the (a) first and (b) second EP encircling. The contours of the stationary states are displayed at the corners of the sketched trajectory. The initial and final states are highlighted and the phase information of the vortex-antivortex at the EP are displayed below. After the first encircling the topological charge of the vortex pair remains unchanged (highlighted in grey and orange). After the second encircling the topological charge of the vortex pair is switched (red).}
  \label{fig:sm2}
\end{figure*}
\end{widetext}

\emph{EP-Encircling} -- The encircling of an EP and the resulting state transitions, including the accumulation of geometric Berry phases, were originally discussed in Ref.~\citep{dembowski2004encircling}. Here, we show that in this system, the orbital process can lead to a change in the topological charge of the vortex-antivortex pair upon reentry into the EP. To this end, the width of the rectangular potential is slightly reduced to $8.7~\mathrm{\upmu m}$ to lift the degeneracy of the two modes. This shifts the tripole mode [Fig. 1(d)] above the dipole mode [Fig. 1(c)] of the system. In addition to the excitation pump $P_1(x,y)$, a second elliptical control pump $P_2(\textbf{r})$ is introduced. The pump profile is thus described by,
\begin{equation}
    P(x,y)=I_1\mathrm{exp}\left(-\frac{x^2+y^2}{\omega^2}\right)^6+I_2\mathrm{exp}\left(-\left|\frac{x^2}{\omega_x^2}+\frac{y^2}{\omega_y^2}\right|\right)^2
\end{equation}
Here $I_1=\in[14;~15]~\mathrm{ps^{-1} \upmu m^{-2}}$ and $\omega=12~\mathrm{\upmu m}$ describe the intensity and width of the excitation pump. The shape of the elliptic pump is defined by its width $\omega_x=5~\mathrm{\upmu m}$ in the x-direction and $\omega_y=2~\mathrm{\upmu m}$ in the y-direction. $I_2=\in[0;~2]~\mathrm{ps^{-1} \upmu m^{-2}}$ describes the intensity of the elliptical pump. Since the saturable gain affects the gain difference of the two modes, the excitation pump is used in the following to control these differences. In addition, the elliptic pump induces a spatially modulated blueshift which, due to its shape, mainly affects the dipole mode of the system. Thus, both energy and gain difference of the modes can be controlled, which allows for an all-optically induced encircling of the EP.
\\
We note that in this case the encircling trajectories are not parallel to the $\Delta E$ and $\Delta\Gamma$ axes, since both pumps cause an individual blue and gain shift to the two modes. Figure \ref{fig:sm2} shows the first and second encircling processes. It shall be emphasized that due to the nonlinearity of the system, the states exemplified at the corners of the encircling process are stationary  before converging to the EP mode. During the first encircling, the dipole mode gathers a $-\pi$ Berry phase, while the tripole mode and hence the EP mode remain unchanged. Only after a second encircling does the tripole mode also collect a geometric Berry phase, leading to the switching of the topological charge of the vortex-antivortex pair.
\emph{Riemann surface Rotation} -- To derive the expression for $\Delta|\psi|^2$ in dependence of $\Delta\Gamma$ we perform the Taylor expansion on the average mode gain and loss ratio $\overline{\Gamma}=\frac{\hbar}{2}\left[\frac{RP}{\gamma_\mathrm{r}+R|\overline{\psi}|^2}-\gamma_\mathrm{c}\right]|\overline{\psi}|^2$ at $|\overline{\psi}|^2=0$ given the approximation of constant average polariton density $|\overline{\psi}|^2$ and pump intensity $P=I_1$. The solution converges for $|\overline{\psi}|^2<\sfrac{\gamma_r}{R}$. We note that this does not limit the strength of nonlinearity that can be observed with this approximation, as $g_c$ can be tuned freely to model strong nonlinear effects. The Taylor expansion of the gain and loss ratio reads
\begin{eqnarray}
        \overline{\Gamma}=\frac{\hbar}{2}\left(\left[\frac{RP}{\gamma_r}-\gamma_c\right]|\overline{\psi}|^2-\frac{PR^2}{\gamma_r^2}|\overline{\psi}|^4+\frac{PR^3}{\gamma_r^3}|\overline{\psi}|^6\mp...\right).\nonumber
\label{eq:taylor}
\end{eqnarray}
From this the mode density difference of the two modes $\upalpha$ and $\upbeta$ as a function of their gain and loss difference can be derived as
\begin{widetext}
\begin{eqnarray}
        \Delta\Gamma &=& \frac{\hbar}{2}\left(\left[\frac{RP}{\gamma_r}-\gamma_c\right]\left(|\overline{\psi_\upalpha}|^2-|\overline{\psi_\upbeta}|^2\right)-\frac{PR^2}{\gamma_r^2}\left(|\overline{\psi_\upalpha}|^4-|\overline{\psi_\upbeta}|^4\right)+\frac{PR^3}{\gamma_r^3}\left(|\overline{\psi_\upalpha}|^6-|\overline{\psi_\upbeta}|^6\right)\mp...\right)\nonumber\\    
        \Leftrightarrow\Delta\Gamma&=&\frac{\hbar}{2}\left(\left[\frac{RP}{\gamma_r}-\gamma_c\right]\Delta|\psi|^2-\frac{PR^2}{\gamma_r^2}\left(|\overline{\psi_\upalpha}|^2+|\overline{\psi_\upbeta}|^2\right)\Delta|\psi|^2+\frac{PR^3}{\gamma_r^3}\left(|\overline{\psi_\upalpha}|^4+|\overline{\psi_\upalpha}|^2|\overline{\psi_\upbeta}|^2+|\overline{\psi_\upbeta}|^4\right)\Delta|\psi|^2\mp...\right)\nonumber\\
        \Leftrightarrow\Delta|\psi|^2[\Delta\Gamma] &=& \frac{2\Delta\Gamma}{\left[\frac{\hbar RP}{\gamma_r}-\hbar\gamma_c\right]-\frac{\hbar PR^2}{\gamma_r^2}\left(|\overline{\psi_\upalpha}|^2+|\overline{\psi_\upbeta}|^2\right)+\frac{\hbar PR^3}{\gamma_r^3}\left(|\overline{\psi_\upalpha}|^4+|\overline{\psi_\upbeta}|^2|\overline{\psi_\upbeta}|^2+|\overline{\psi_\upbeta}|^4\right)\mp...}\nonumber\\
        \Rightarrow\Delta|\psi|^2[\Delta\Gamma]&\approx& \frac{2\Delta\Gamma}{\left[\frac{\hbar RP}{\gamma_r}-\hbar\gamma_c\right]-\frac{\hbar PR^2}{\gamma_r^2}\left(|\overline{\psi_\upalpha}|^2+|\overline{\psi_\upbeta}|^2\right)+\frac{\hbar PR^3}{\gamma_r^3}\left(|\overline{\psi_\upalpha}|^2+|\overline{\psi_\upbeta}|^2\right)^2\mp...}\nonumber\\
        \Rightarrow\Delta|\psi|^2[\Delta\Gamma]&=& \frac{2\Delta\Gamma}{\left[\frac{\hbar PR}{\gamma_\mathrm{r}}-\hbar\gamma_\mathrm{c}\right]+\sum_{m=2}^{\infty}\left(-1\right)^{m-1}\frac{\hbar PR^m}{\gamma_\mathrm{r}^m}\left(|\psi_\mathrm{tot}|^2\right)^{m-1}}.
\label{eq:dgamma}
\end{eqnarray}
\end{widetext}
In the second step, the terms of the form $\left(|\overline{\psi_\upalpha}|^2\right)^m-\left(|\overline{\psi_\upbeta}|^2\right)^m$ are factorized with respect to the density difference $\Delta|\psi|^2$, according to
\begin{equation}
    x^m-y^m=(x-y)\sum_{l=0}^{m-1}x^ly^{m-1-l}.   
\end{equation}
In the last step, the factors of the $m$-th order terms are approximated by the total density $\psi_\mathrm{tot}$=$|\psi_\upalpha|^2+|\psi_\upbeta|^2$ to the $(m-1)$-th power. By this approximation, the denominator is always larger than its exact value for the parameter range bounded by the Taylor expansion. Accordingly, this approximation describes a lower bound for the mode density difference under the independent variation of $\Delta\Gamma$. We emphasize that this approximation is exact for the first two terms in the denominator and the strengths of the higher-order terms decrease by one order of magnitude for each order $m$. Thus, we conclude that this approximation does not significantly affect our results. We note, that the denominator vanishes when the saturable gain compensates the linear gain. For larger densities the direction of the rotation is inverted. In regard to the approximations described above, we collect our results away from this divergence.\\

To determine the rotation and blueshift of the Riemann surface induced by the excitation reservoir in a polariton system we derive the expression for $\Delta n[\Delta |\psi|^2]$ and $n_\mathrm{tot}[|\psi_\mathrm{tot}|^2]$ from the Taylor expansion of $\Gamma$ given above and the relation of $n$ and $\Gamma$. The resulting blueshift induced on the Riemann surface is defined by
\begin{equation}
    n_\mathrm{tot}=\left[\frac{2P}{\gamma_\mathrm{r}}-\frac{2\gamma_\mathrm{c}}{R}\right]+\sum_{m=1}^{\infty}(-1)^m\frac{PR^m}{\gamma_\mathrm{r}^{m+1}}\left(|\psi_\mathrm{tot}|^2\right)^m.
\end{equation}
The rotation induced by the excitation reservoir is given by
\begin{equation}
    \Delta n= \sum_{m=1}^{\infty}(-1)^m\frac{PR^m}{\gamma_\mathrm{r}^{m+1}}\left(|\psi_\mathrm{tot}|^2\right)^{m-1}\Delta|\psi|^2.
\end{equation}
From the given expression it can be seen that the rotation induced by the reservoir blueshift rotates the Riemann surface even further.
\end{document}